\documentclass[12pt]{iopart}

\usepackage[dvips]{graphicx}
\usepackage{mathtext}       
\usepackage[english]{babel}
\usepackage{citehack}       

\newcommand{\lb}{\left(}
\newcommand{\rb}{\right)}
\newcommand{\pd}{\partial}

\begin{document}

\title[]
{Similarity Solution of the 3-phase Stefan Problem for Alloys with Arbitrary Temperature Dependent Properties.}
\author{Evgeniy N.~Kondrashov 
}
\address{JSC VSMPO-AVISMA Corporation, Titanium Alloy Ingot Group, R\&D Center,\\
Parkovaya str.~1, Verkhnyaya Salda, 624760, Sverdlovsk region, Russian Federation}
\ead{evgeniy.kondrashov@vsmpo.ru}

\begin{abstract}

In this paper a 3-phase Stefan problem solution method for 1D semi-infinity alloy is developed. The problem is first solved for full enthalpy
of the system and then the thermal diffusivity has been eliminated from the divergence operator by Kirchoff transformation. Moreover, we introduce
a similarity independent variable $\eta=x^{2}/\tau$ and original problem transforms to ordinary differential equations (ODE) for each
phase separately. These ODEs have Dirichlet's type boundary conditions.

Furthermore, we give an solution example of these equations for a simple case.

\end{abstract}







\section{Problem and Full Enthalpy Equation. \label{enthalpy_eq}}

We formulate the problem for the temperature field $T(x, \tau)$
(where $x$ is a coordinate and $\tau$ is the time) in following form.
We have a 1D semi-infinite liquid alloy array at the temperature $T_{init} > T_{l}$,
where $T_{l}$ is the liquidus temperature of alloy.
The alloy array is located at $x \geq 0$.
In the initial time $\tau = 0$ the temperature at the point $x=0$
instantly drops to $T_{out} < T_{s}$, where $T_{s}$ is the solidus
temperature of alloy.
Thereafter we have a consistent solidification process from left to right.
We need to find $T(x,\tau)$ and solidus $X_{s}(\tau)$ and liquidus
$X_{l}(\tau)$ fronts locations.
In any moment (except initial moment) there {\it exist} 3 phases:
\begin{itemize}
\item		Solid for $0 \leq x \leq X_{s}(\tau)$.
\item		Solid-liquid (mushy zone) for $X_{s}(\tau) \leq x \leq X_{l}(\tau)$.
\item		Liquid for $x \geq X_{l}(\tau)$.
\end{itemize}
In this paper, we suppose $\rho$ is constant and the same in each phase. Heat capacity $C(T)$
and thermal conductivity $\kappa(T)$ are arbitrary single-valued functions of temperature $T$.
Mathematical the problem can be expressed as follows
\begin{eqnarray}
\rho C(T) \frac{\pd T}{\pd \tau} = div \lb \kappa(T) grad T \rb + \rho L \frac{\pd \lambda}{\pd \tau}, \label{T_eq} \\
T(x, \tau = 0) = T_{init}, \\
T(0, \tau) = T_{out}, \quad \tau > 0, \\
T(\infty, \tau) = T_{init}, \quad \tau \geq 0,
\end{eqnarray}
where $L$ is latent heat, $\lambda$ is liquid fraction. We introduce the full enthalpy as
\begin{equation}
H(T) = \rho \int\limits_{0}^{T} C(t) dt + \rho L \lambda(T)
\label{h_t}
\end{equation}
then Eq.~(\ref{T_eq}) can be rewritten in the form
\begin{equation}
\frac{\pd H}{\pd \tau} = div \lb \kappa(T) grad T \rb.
\label{HT_eq}
\end{equation}
With regard to thermodynamical identity
\begin{equation}
grad T = \frac{grad H}{\frac{dH}{dT}},
\end{equation}
we rewrite Eq.~(\ref{HT_eq}) with initial and boundary conditions as

\begin{eqnarray}
\frac{\pd H}{\pd \tau} = div \lb \alpha(H) grad H \rb, \label{H_eq} \\
H(x, \tau = 0) = H_{init} = H(T_{init}), \\
H(0, \tau) = H_{out} = H(T_{out}), \quad \tau > 0 \\
H(\infty, \tau) = H_{init} = H(T_{init}), \quad \tau \geq 0,
\end{eqnarray}
where $\alpha$ is thermal diffusivity. The temperature dependence of the thermal diffusivity is
defined by

\begin{equation}
\tilde{\alpha}(T) = \frac{\kappa(T)}{\frac{dH(T)}{dT}}.
\end{equation}

The function (\ref{h_t}) can be invert to $T=T(H)$, therefore in (\ref{H_eq}) we write
$\tilde{\alpha}(T) = \tilde{\alpha}(T(H)) = \alpha(H)$.
It should be noted that the function $\alpha(H)$ has discontinuities at the $X_{s}$ and $X_{l}$
\begin{equation}
\left. \alpha(H) \right|_{H = H_s - 0} = \alpha_{s},
\qquad
\left. \alpha(H) \right|_{H = H_s + 0} = \alpha_{ms},
\end{equation}

\begin{equation}
\left. \alpha(H) \right|_{H = H_l - 0} = \alpha_{ml},
\qquad
\left. \alpha(H) \right|_{H = H_l + 0} = \alpha_{l},
\end{equation}
where we denote

\begin{equation}
H_{s} = H(T_{s}),
\qquad
H_{l} = H(T_{l}).
\end{equation}
Now, we use the Kirchoff transformation to eliminate $\alpha$ from below the divergence operator.
We introduce the new function $Z(x, \tau)$ as

\begin{equation}
Z(H) = \int\limits_{0}^{H} \alpha(h)dh.
\label{Z_H}
\end{equation}
For $Z(x, \tau)$ the original problem is given by

\begin{eqnarray}
\frac{1}{\beta(Z)} \frac{\pd Z}{\pd \tau} = \Delta Z, \label{Z_eq} \\
Z(x, \tau = 0) = Z_{init} = Z(H_{init}), \\
Z(0, \tau) = Z_{out} = Z(H_{out}), \quad \tau > 0 \\
Z(\infty, \tau) = Z_{init} = Z(H_{init}), \quad \tau \geq 0,
\end{eqnarray}
where $\Delta$ is the Laplace operator, and $\alpha(H) = \alpha(H(Z)) = \beta(Z)$,
since we can invert Eq.~(\ref{Z_H}) into $H=H(Z)$.

\section{The Transformation of the Problem to ODE. \label{ODE}}

Equation (\ref{Z_eq}) can be transformed to a nonlinear ordinary differential equation (ODE).
We introduce new independent similarity variable

\begin{equation}
\eta = \frac{x^{2}}{\tau},
\end{equation}
then we can write

\begin{eqnarray}
\frac{\pd Z}{\pd \tau} = - \frac{\eta}{2\tau}\frac{dZ}{d\eta},
\qquad
\frac{\pd Z}{\pd x} = \frac{1}{\sqrt{\tau}}\frac{dZ}{d\eta},
\qquad
\frac{\pd^{2} Z}{\pd x^{2}} = \frac{1}{\tau} \frac{d^{2}Z}{d\eta^{2}}.
\label{transf_xt_eta}
\end{eqnarray}
The Eq.~(\ref{Z_eq}) can be written as
\begin{equation}
\frac{d^{2}Z}{d\eta^{2}} = - \frac{\eta}{2\beta(Z)} \frac{dZ}{d\eta}.
\label{Z_ODE}
\end{equation}
The initial and boundary conditions are

\begin{equation}
Z(\eta = 0) = Z_{out},
\qquad
Z(\eta \rightarrow \infty) = Z_{init}.
\end{equation}
Now, we suppose that the solidus/liquidus fronts positions are varying as

\begin{equation}
X_{s}(\tau) = k_{s} \sqrt{\tau},
\qquad
X_{l}(\tau) = k_{l} \sqrt{\tau},
\end{equation}
where $k_{s}$ and $k_{l}$ are constants.
This dependence is the standard one for the 1D Dirichlet heat transfer problem with phase change, which has many
analytical and numerical solutions (see the references in \cite{ken-towards}). Now we can divide the equation for each phase.
Let $\beta(Z)$ be

\begin{eqnarray}
\beta(Z) = \cases{
\beta_{s}(Z)	& 	for	$Z \in [0, Z_{s})$, \\
\beta_{m}(Z)	&	for	$Z \in [Z_{s}, Z_{l}]$, \\
\beta_{l}(Z)	&	for $Z \in (Z_{l}, \infty)$,
}
\end{eqnarray}
where $Z_{s, l} = Z(H_{s, l})$. For each phase the equations have the uniform form

\begin{equation}
\frac{d^{2}Z}{d\eta^{2}} = - \frac{\eta}{2\beta_{j}(Z)} \frac{dZ}{d\eta},
\qquad
j = s, m, l.
\label{Z_j_ODE}
\end{equation}
The boundary conditions are

\begin{eqnarray}
Z(\eta = 0) = Z_{out}, \qquad Z(\eta = k_{s}) = Z_{s}, \quad solid \; phase \; (s) \\
Z(\eta = k_{s}) = Z_{s}, \qquad Z(\eta = k_{l}) = Z_{l}, \quad mushy \; phase \; (m) \\
Z(\eta = k_{l}) = Z_{l}, \qquad Z(\eta \rightarrow \infty) = Z_{init}, \quad liquid \; phase \; (l).
\end{eqnarray}
Moreover, we have to define heat flux conditions on the interfaces between the phases. For $Z$ these conditions are

\begin{eqnarray}
\left. \frac{\pd Z}{\pd x} \right|_{x=X_{s}-0} = \left. \frac{\pd Z}{\pd x} \right|_{x=X_{s}+0} + \rho \lambda_{0} L \frac{dX_{s}(\tau)}{d\tau}, \label{Z1} \\
\left. \frac{\pd Z}{\pd x} \right|_{x=X_{l}-0} = \left. \frac{\pd Z}{\pd x} \right|_{x=X_{l}+0}, \label{Z2}
\end{eqnarray}
where $\lambda_{0}$ -- eutectic liquid fraction (for eutectic alloys). For the enthalpy we can write

\begin{eqnarray}
\left. \alpha_{s} \frac{\pd H}{\pd x} \right|_{x=X_{s}-0} = \left. \alpha_{ms} \frac{\pd H}{\pd x} \right|_{x=X_{s}+0} + \rho \lambda_{0} L \frac{dX_{s}(\tau)}{d\tau}, \label{H1} \\
\left. \alpha_{ml} \frac{\pd H}{\pd x} \right|_{x=X_{l}-0} = \left. \alpha_{l} \frac{\pd H}{\pd x} \right|_{x=X_{l}+0}, \label{H2}
\end{eqnarray}
The unknown parameters $k_{s}$ and $k_{l}$ will be determined from either (\ref{Z1})-(\ref{Z2}) or (\ref{H1})-(\ref{H2}).

\section{Example. \label{examples}}

The goal of this section is to give an example of the 3-phase Stefan problem solution. We will examine the problem
with a constant thermal diffusivities in each phase $\alpha_{s} \ne \alpha_{m} \ne \alpha_{l}$.
Analogous problems were studied in \cite{ken-towards, ken-jet-density}.
In this case we have $\beta_{j} = \alpha_{j}$ for $j=s,m,l$.
The general solution of the Eq.~(\ref{Z_j_ODE}) is given by

\begin{equation}
Z_{j}(\eta) = C_{1} + C_{2} \sqrt{\pi \alpha_{j}} erf\lb \frac{\eta}{2\sqrt{\alpha_{j}}} \rb,
\end{equation}
where $C_{1}$ and $C_{2}$ are arbitrary constants.
Thus, it is to easy obtain expressions for the enthalpies ($H_{j} = Z_{j}/\alpha_{j}$) in each phase

\begin{equation}
\fl H(x,t) = H_{out} + (H_{s} - H_{out})
\frac{erf\left( \frac{x}{2\sqrt{\alpha_{s} t}} \right)}{erf \lb \frac{k_{s}}{2\sqrt{\alpha_{s}}} \rb },
\qquad
x \in [0, X_{s}),
\label{h_solution_s}
\end{equation}

\begin{equation}
\fl H(x,t) = \frac{ (H_{l}-H_{s}) erf \lb \frac{x}{2\sqrt{\alpha_{m} t}} \rb
+ H_{s} erf \lb \frac{k_{l}}{2\sqrt{\alpha_{m}}} \rb -
H_{l} erf \lb \frac{k_{s}}{2\sqrt{\alpha_{m}}} \rb     }
{  erf \lb \frac{k_{l}}{2\sqrt{\alpha_{m}}} \rb  -
erf \lb \frac{k_{s}}{2\sqrt{\alpha_{m}}} \rb  },
\quad
x \in [X_{s}, X_{l}],
\label{h_solution_sl}
\end{equation}

\begin{equation}
\fl H(x,t) = 
H_{init} - (H_{init} - H_{l}) 
\frac{  erfc \lb \frac{x}{2\sqrt{\alpha_{l}t}} \rb  }
{  erfc \lb \frac{k_{l}}{2\sqrt{\alpha_{l}}} \rb   },
\quad
x \in (X_{l}, \infty),
\label{h_solution_l}
\end{equation}
The equation system (\ref{H1})-(\ref{H2}) for the determination of $k_{s}$ and $k_{l}$ can be written as

\begin{equation}
\fl \frac{ \sqrt{\alpha_{s}}(H_{s} - H_{out}) exp\lb -\frac{k_{s}^{2}}{4\alpha_{s}} \rb }
{erf\lb \frac{k_{s}}{2\sqrt{\alpha_{s}}} \rb}
-
\frac{ \sqrt{\alpha_{m}} (H_{l} - H_{s}) exp\lb - \frac{k_{s}^{2}}{4\alpha_{m}} \rb }
{erf\lb \frac{k_{l}}{2\sqrt{\alpha_{m}}} \rb - erf\lb \frac{k_{s}}{2\sqrt{\alpha_{m}}} \rb }
= \frac{\sqrt{\pi}}{2} \rho \lambda_{0} L k_{s},
\label{cond-s-real}
\end{equation}

\begin{equation}
\fl \frac{ \sqrt{\alpha_{m}} (H_{l} - H_{s}) exp\lb - \frac{k_{l}^{2}}{4\alpha_{m}} \rb }
{erf\lb \frac{k_{l}}{2\sqrt{\alpha_{m}}} \rb - erf\lb \frac{k_{s}}{2\sqrt{\alpha_{m}}} \rb }
-
\frac{ \sqrt{\alpha_{l}}(H_{init} - H_{l}) exp\lb -\frac{k_{l}^{2}}{4\alpha_{l}} \rb }
{erfc\lb \frac{k_{l}}{2\sqrt{\alpha_{l}}} \rb}
= 0
\label{cond-l-real}
\end{equation}
This system is easy to solve by numerical methods (see \cite{ken-towards} and references therein).


\end{document}